\documentclass[aps,%
pre,%
twocolumn,%
showpacs,%
floatfix,%
preprintnumbers,%
superscriptaddress]{revtex4}
\usepackage{epsfig} 
\newcommand{\IGN}[1]{}
\begin{document}
\preprint{to appear in Phys. Rev. E (2004)}
\author{Jens Christian Claussen}
\affiliation{Institut f\"ur Theoretische Physik und Astrophysik, Universit\"at Kiel, Leibnizstra{\ss}e 15, 24098 Kiel, Germany}
\author{Jan Nagler}
\affiliation{Institut f\"ur Theoretische Physik, Universit\"at Bremen, Otto-Hahn-Allee, 28334 Bremen, Germany}
\author{Heinz Georg Schuster}
\affiliation{Institut f\"ur Theoretische Physik und Astrophysik, Universit\"at Kiel, Leibnizstra{\ss}e 15, 24098 Kiel, Germany}
\title{Sierpinski signal generates $1/f^\alpha$ spectra}
\date{August 14, 2003; revised April 28, 2004}
\newcommand{\mod}{{\rm{}mod}}
\begin{abstract}
We investigate the row sum of the binary
pattern generated by the Sierpinski automaton:
Interpreted as a time series
we calculate the power spectrum of this Sierpinski signal
analytically and obtain a unique rugged fine structure
with underlying power law decay with
an exponent of approximately 1.15.
Despite the simplicity of the model,
it can serve as a model for $1/f^\alpha$ spectra
in a certain class of experimental and natural systems
like catalytic reactions and mollusc patterns.
\end{abstract}
\maketitle
The phenomenon of
 $1/f^\alpha$ noise is found in a widespread variety
of systems \cite{jensen,hownatureworks,davidsenschuster02}.
Usually a noise signal is said to be
$1/f^\alpha$ (or $1/f$)
if its spectrum
follows over some decades
a power law
$S(\omega) \sim \omega^{-\alpha}$
 with an exponent near to one.
Despite $1/f$ noise is being known for a long time, up
to now there is no general explanation for universal mechanisms
(if they exist).
However, on a very general
level it is believed
that complex systems
\cite{cas}
are able to generate $1/f$ noise.
While real complex systems are usually not exactly solvable, extremely
simplified models are under investigation. One prominent example
is the sandpile-sketching Bak-Tang-Wiesenfeld model \cite{btw}, a
two-dimensional cellular automaton, which itself does however not
reproduce $1/f$ sufficiently.

\begin{figure}[htbp]
\noindent
 \epsfig{file=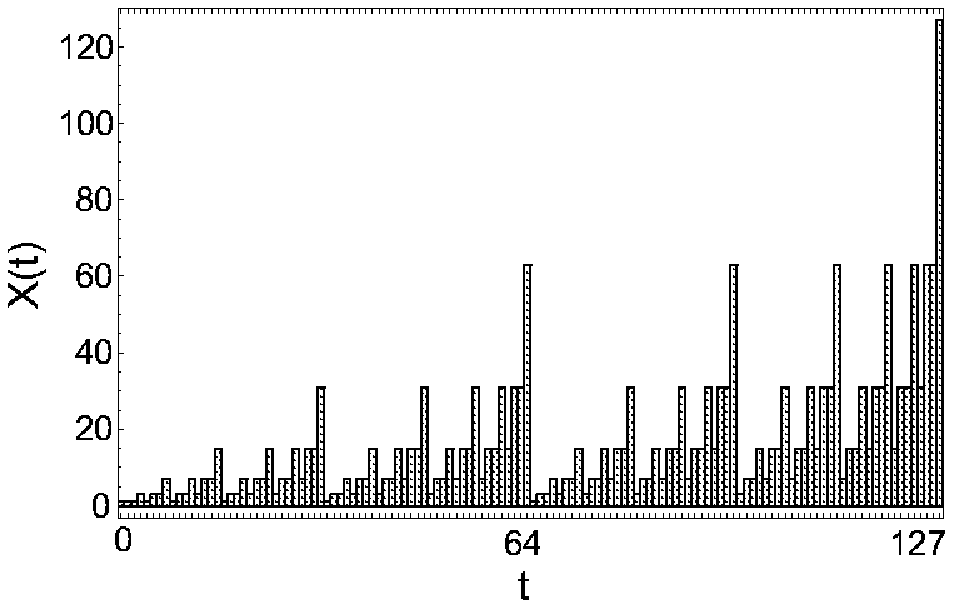,angle=0,height=5cm}
\noindent
\caption{The self-similar {\sl Sierpinski signal} $X(t)$ for $T=128$.
\label{fig_sierp0}}
\end{figure}

In this work we study an even more simplified model introduced in
1984 by Wolfram \cite{wolfram84}, which is known to be able to
exhibit complex behavior: the Sierpinski automaton. Looking not at
the generated fractal Sierpinski gasket itself, but on the {\sl
row sum},
corresponding to the total {\em (in-)activity} of the whole
system, we have a signal as shown in fig{.\ref{fig_sierp0}} with
increasing mean and increasing spatial size of the corresponding
system; thus every physical realization of the system will be
finite size-limited.

Despite on a first glance being a theoretical toy model only,
Sierpinski patterns have been found in nature. Detailed models
have explained mollusc patterns by reaction-diffusion models and
cellular automata \cite{meinhardtbook,meinhardt97}; Sierpinski
patterns also occur in kinkbreeding dynamics \cite{chate99} and
have been observed in catalysis \cite{otterstedt98}. This
phenomenon occurs generically for suitable parameter choices in
four standard types of nonlinear spatiotemporal dynamics including
the Bonhoeffer-van der Pol and the complex Ginzburg-Landau
equation \cite{hayase}.

Consequently,
catalytic processes can exhibit
similar time-signals as the Sierpinski sum signal.
A comparison of the reaction rate of a catalytic process
with the Sierpinski sum signal
 $X(t)=\sum x_i(t)$
has been given in \cite{dress84plath,liauw96}.
A single state $x_i(t)$ at a time $t$ can be interpreted to
indicate the activity of a local catalysis process, {i.e.} {\em
reaction (activity)} when $x_i(t)=0$ and {\em no reaction
(inactivity)} when $x_i(t)=1$.
The authors \cite{dress84plath} observe a
qualitative similarity between the experimental and theoretical
time series.
Due to dominating finite size effects
a $1/f^{\alpha}$ spectra (or long-time correlations)
could not be identified in the
spectrum of the experimental data
\cite{plathnaglerprivcomm}.
As models of chemical reactions,
cellular automata have been studied widely
\cite{gerhardtschusterphysicad}, explaining spiral waves and
pattern formation in chemical reactions. CO-oxidation on Pt(110)
and its control by global delayed feedback has been studied and
compared to models \cite{kimbertrametal}, including the occurrence
of patterns similar to Sierpinski structures in the intermittent
turbulent phase.
Recently,  $1/f^\alpha$ spectra have been measured
directly in a chemical reaction \cite{claycomb01} by a SQUID
setup that allows for much higher resolution in time, space and
signal-to-noise ratio than the direct gravimetric
measurement of the reaction rate in \cite{dress84plath}.
The power law extends over more than two decades,
indicating the spatiotemporal dynamics of the catalytic
reaction exhibits avalanches on all sizes
and self-organized critical behavior \cite{claycomb01}.
Interestingly, the Sierpinski gasket was more recently found in a video feedback system \cite{nature}.
Apart from observation of the Sierpinski pattern itself, its
geometry has been used widely,
{e.g.}, for sandpile dynamics
\cite{sandpile} and measurements of magnetoresistance on fractal
wire networks \cite{wire}.

\paragraph*{Definition of the model:}
The dynamics of the so-called {\sl Sierpinski  automaton} is
related to the generation law of the Pascal triangle. This pattern
can be generated by the following simple one-dimensional cellular
automaton: We consider a linear array of sites (or spins) $x_i(t)$
which can take the values 0 or 1 at discrete time steps $t$.
We restrict ourselves to the
special initial condition,
that for $t=0$
only one spin is different from all others:
\vspace*{-.4ex}
\begin{eqnarray}
x_0(0)=1
~~~~~~
{\rm and}
~~~~~~
\forall_{i\neq{}0} ~~~ x_i(0)=0.
\end{eqnarray}
\vspace*{-.3ex}
\noindent
The dynamics is defined  by the following next-neighbor interaction:
\vspace*{-.4ex}
\begin{eqnarray}
x_i(t+1)=[x_{i-1}(t)+x_{i+1}(t)] ~~ \mod ~~ 2,
\end{eqnarray}
\vspace*{-.3ex}
$ i\in [-\infty,
\infty ]$,
$x_i(t) \in \{0,1\}$ at discrete time  $t$.
In the context of catalytic processes \cite{otterstedt98,dress84plath,liauw96}
a simplified chemical interpretation of this rule reads:
A catalytic process is stopped
when too less (i.e. no) or too many (i.e. 2) neighbor sites are active.
A catalytic process is initiated (or continued) when
only one neighbor site is active.
This can origin from
a minimal catalysis temperature combined with
a local self-limiting reaction rate.
 \clearpage
\noindent
\IGN{The spatiotemporal evolution 
is 
equivalent to 
the rows
of 
the Pascal triangle after entries have been taken modulo~2,
}
The spatiotemporal evolution
is obtained from the rows of the Pascal triangle by
applying modulo 2 elementwise,
\vspace{-.4ex}
\begin{eqnarray}
\nonumber
\begin{array}{ccccccccccccccccccc}
&&&&1
 &&&&&&
 &&&&1\\
&&&1&&1
 &&&&&
&&&1&&1\\[-2ex]
&&1&&2&&1
&&& \stackrel{\mbox{\normalsize\rm modulo 2}}{\mbox{\Large$\longrightarrow$}} &
&&1&&0&&1\\
&1&&3&&3&&1
&&&&1&&1&&1&&1\\
1&&4&&6&&4&&1
&
&1&&0&&0&&0&&1
\end{array}
\end{eqnarray}

\noindent
which is also known as the Sierpinski gasket.
The Sierpinski gasket is a well-known self-similar
structure (with point dimension $\ln{}3/\ln{}2$
in $x$ for $t\to{}\infty$)
that is obtained by twofold replication
of the first four rows to the subsequent four rows,
and iteration of this process with the whole triangle.

Instead of considering the fractal pattern itself, we look at a
scalar observable that can be compared to experimental time
series. Before considering the spectrum, we briefly sketch a
direct solution \cite{kummer1852} and illustrate the analogy to a
formal language approach. The row sum (or total (in-)activity)
over space at time $t$, defined by
\begin{eqnarray}
X(t):=\sum_i x_i(t),
\end{eqnarray}
is referred to as {\sl Sierpinski signal}.
The Sierpinski
automaton rule then
generates a time series $X(t)$
(fig.~\ref{fig_sierp0})
starting from $t=0$ with
\begin{eqnarray}
1,~~2,~~2,4,~~2,4,4,8,~~2,4,4,8,4,8,8,16,~~\ldots
\label{eq:zeitreihe}
\end{eqnarray}
%
%
%
%
\noindent
\noindent
Being interested in an analytic expression for
(\ref{eq:zeitreihe}), we first note that
$X(t)$ can be generated
up to $t=2^N-1$
from the {\sl start sequence}
$u_0=(1)$ by $N$ iterations of the
sequence replication rule
\begin{eqnarray}
u_n \rightarrow
u_{n+1}=(u_n, 2 \cdot u_n).
\label{eq_formal}
\end{eqnarray}
\\
Obviously $X(t)$ takes only powers of 2,
so we
consider $(\ln X(t))/\ln 2$, which starts as
\begin{eqnarray}
0,~~1,~~1,2,~~1,2,2,3,~~1,2,2,3,2,3,3,4,~~\ldots
\label{eq:zeitreihe_l}
\end{eqnarray}
Strikingly, this appears to be the number of ones in
\begin{eqnarray}
\nonumber
0,01,10,11,100,101,110,111,1000,1001,\\1010,1011,1100,1101,1111,\ldots,
\end{eqnarray}
{i.e.} the binary decomposition
of the time variable $t$ starting with $t=0$.
Therefore the observable is given by
\begin{eqnarray}
X(t)=2^{\displaystyle {\rm CrossSum} ( {\rm Binary} (t))},
\label{binrow}
\end{eqnarray}
which is no longer recursively defined.
This analytic solution has already been known
in 1852 by Kummer \cite{kummer1852}
in a number-theoretic context.

A convenient closed expression can be obtained by expressing time
by a number of {\em spins}
\begin{eqnarray}
t = \sum_{j=0}^{N-1} \sigma_j 2^j
\quad\quad
\sigma \in \{0,1\}.
\end{eqnarray}
and expressing $X(\{\sigma_i\})$ from the same configuration as
\begin{eqnarray}
X(t) = X \!\Big(\! \sum_{j=0}^{N-1} \sigma_j 2^j \!\Big)\!
= 2^{  \sum_{j=0}^{N-1} \sigma_j }.
\;\;\;
\end{eqnarray}
By these expressions for $t(\{\sigma_i\})$ and $X(\{\sigma_i\})$,
we have a parametric expression parameterized by a set of $N$
{\em spins} for all $X(t)$ with times up to $t=2^N-1$.

\begin{figure}[htbp]
\noindent
 \centerline{\epsfig{file=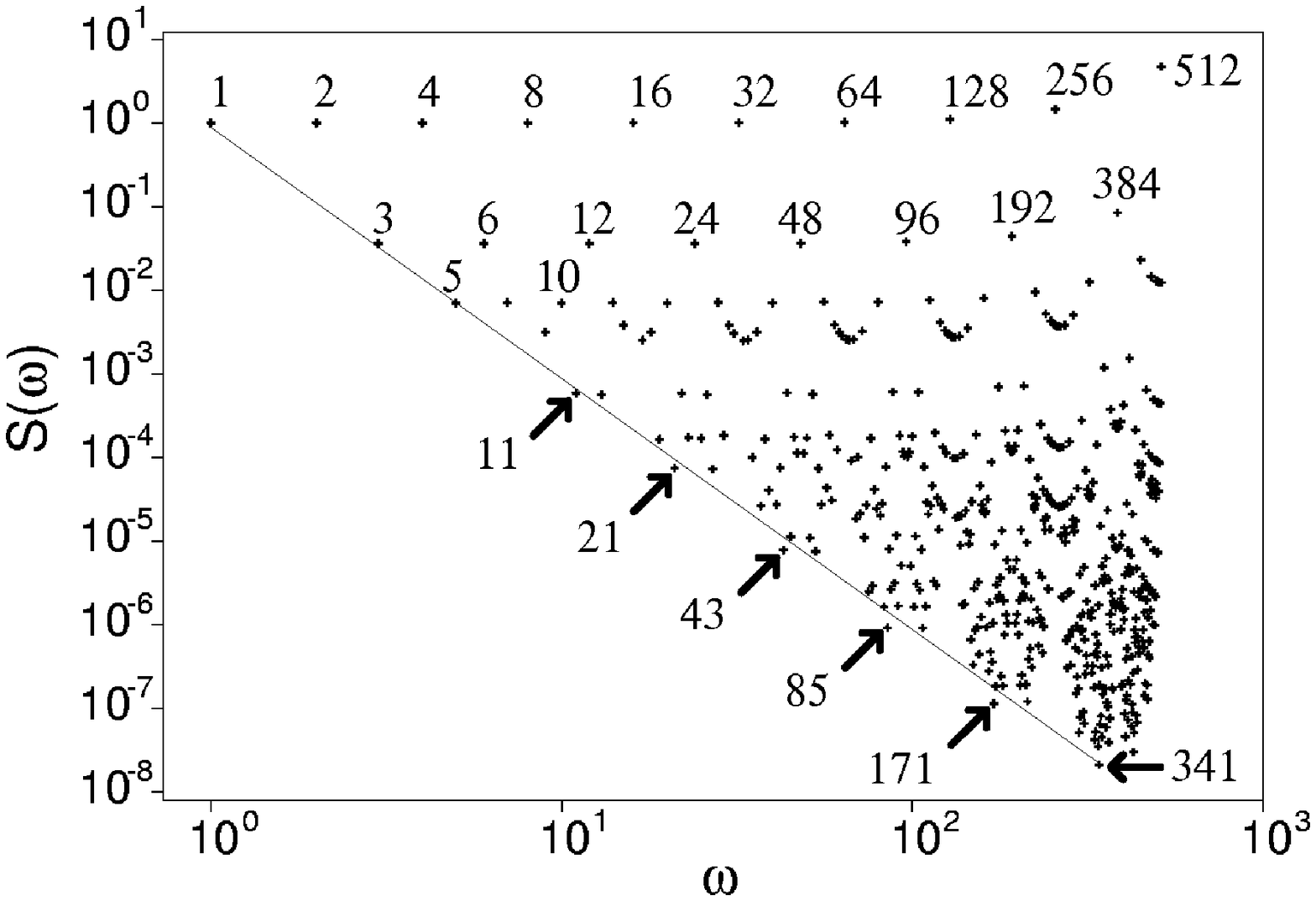,angle=0,width=8.5cm}}
\noindent
\caption{Power spectrum of $X(t)$ for $T=1024$ time steps up to the Nyquist frequency
of $T/2$
(from FFT).
The lower {\em Envelope} is constituted by
$\omega(k)= \lfloor \frac{2^k+1}{3} \rfloor, k\in\{2,3,\ldots\}.$
\label{fig_strukgesamt}}
%
 \centerline{\epsfig{file=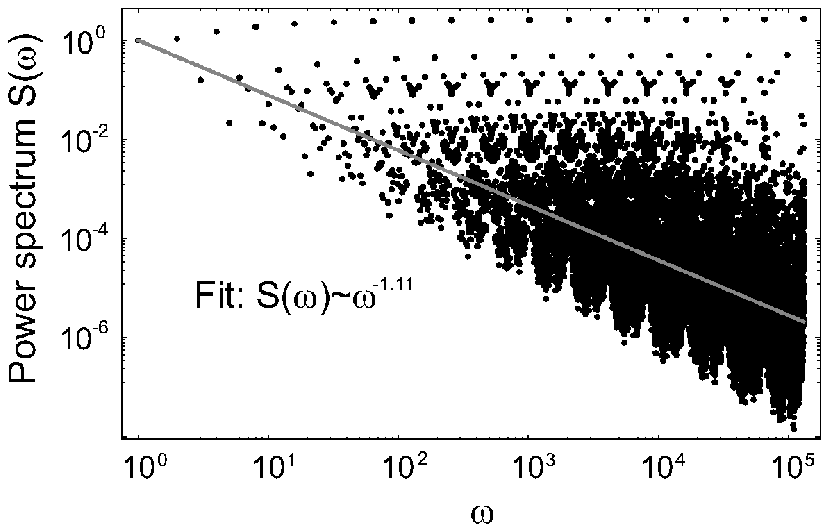,angle=0,width=8.5cm}}
\caption{Power spectrum  of the time signal $Z(t)$ up to $T/8$ for
$T=2^{20}$ time steps (from FFT) and least-square-fit $\omega^{-1.11}$.
 \label{fig_fit}}
 \centerline{\epsfig{file=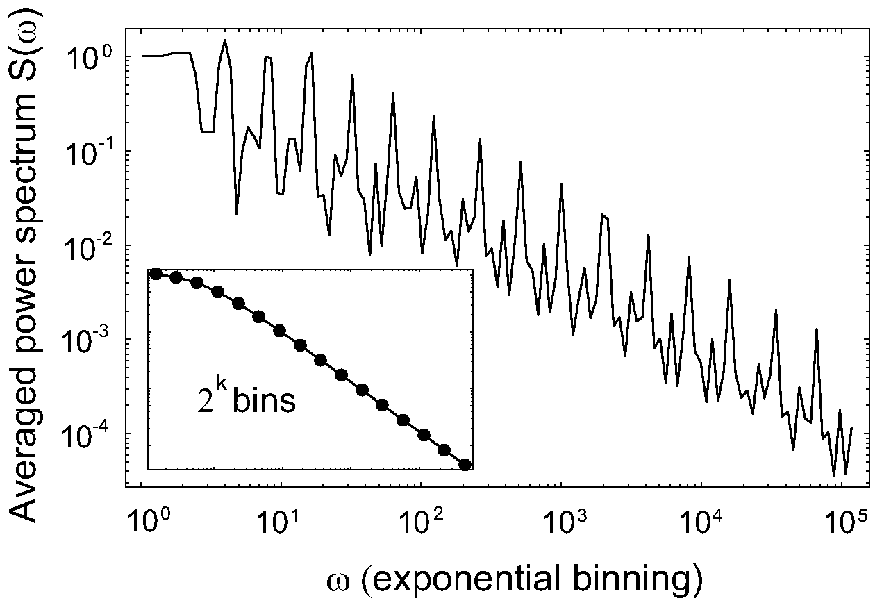,angle=0,width=8.5cm}}
\caption{Averaged power spectrum of $Z(t)$ up to $T/8$ for $T=2^{20}$
using (incommensurable) $1.1^k$-bins, i.e.~
the $k$-th interval is  defined  by $[\lceil 1.1^k\rceil,\lceil
1.1^{k+1} \rceil]$
where the brackets $\lceil \rceil$ denote rounded integer values.
The inset shows the same spectrum,
averaged using $2^k$-bins, i.e.{}
the $k$-th interval is  defined  by $[2^k,2^{k+1}-1]$.
Both correspond to a constant $\delta\omega/\omega$ ratio.
 \label{fig_av}}
 \end{figure}

\paragraph*{Spectrum of the Sierpinski signal:}
The periodogram
$X(\omega)$
of the time signal now is calculated analytically.
The binary time decomposition allows a
Fourier transformation of $X(t)$ fairly direct
from the definition,
\begin{eqnarray}\label{xom}
X(\omega)&=& \sum_{t=0}^{2^N-1} {\rm{}e}^{{\rm{}i}\omega t} X(t)
  =\sum_{\sigma_0}\!\cdots\!\!\sum_{\sigma_{N-1}}
  {\rm{}e}^{{\rm{}i}\omega t(\{\sigma_i\})} X(t(\{\sigma_i\})) \nonumber\\
&=& \sum_{\sigma_0}\!\cdots\!\!\sum_{\sigma_{N-1}} \prod_{j=0}^{N-1}
   \exp\left(  \sigma_j ({\rm{}i} \omega  2^j + \ln 2) \right) \nonumber\\
&=& \prod_{j=0}^{N-1} \sum_{\sigma_j} \exp\left(  \sigma_j
   ({\rm{}i} \omega  2^j + \ln 2) \right) \nonumber\\
&=& \prod_{j=0}^{N-1}
   \left( 1+ \exp ({\rm{}i} \omega  2^j + \ln 2) \right),
\end{eqnarray}
where all sums over $\sigma_j$ are taken over
the two possible values $\sigma_j=0$ and $\sigma_j=1$.
We now calculate the
periodogram's power spectrum, i.~e. $|X(\omega)|^2$.
The absolute value of $X(\omega)$ simplifies
to a trigonometric product which the logarithm converts into a sum,
\begin{eqnarray}
\ln |X(\omega)|^2 = \sum_{j=0}^{N-1} \ln [5+4 \cos(\omega 2^j)],
\label{n1}
\end{eqnarray}
showing a rugged fine structure as shown in fig.~\ref{fig_strukgesamt}.
A rough estimate of the sum in eq. (\ref{n1}) is obtained
approximating the sum by an integral ($y:=\omega 2^j$),
\begin{eqnarray}
\ln |X(\omega)|^2 & \approx& \int_{0}^{N-1} \ln [5+4 \cos(\omega 2^j)]
{\rm d}j
\label{n2}
\\
\label{bothints}
&=& \frac{1}{\ln 2}\int_{\omega}^{\omega
2^{N-1}} \frac{\ln [5+4 \cos y]}{y} {\rm d}y.
\end{eqnarray}
      \clearpage
\noindent
As $\ln(5+4x) \approx \ln(5)
+\frac{4}{5}x$ for $|x|\ll 1$,
we obtain
\begin{eqnarray}
\nonumber
\ln |X(\omega)|^2 \approx \frac{\ln 5}{\ln 2}
\!
\int_{\omega}^{\omega
2^{N-1}}
\!\!
 \frac{ {\rm d}y}{y} +
                           \frac{4}{5\ln 2}
\!
\int_{\omega}^{\omega
2^{N-1}}
\!\!
               \frac{\cos(y)}{y} {\rm d}y.
\end{eqnarray}
\normalsize
\noindent
The integral over the integral cosine
is nearly
independent
of the upper boundary for high values of the boundary. Thus, we
can substitute the upper boundary $\omega 2^{N-1}$ by some
$N$-dependent constant, say $c_N\gg 1$. Finally, substituting the
cosine by one yields immediately a rough approximation of eq. (\ref{n1}):
\begin{eqnarray}\label{result}
|X(\omega)|^2  \approx c_N'
\omega^{-
{4}/{(5\ln 2)}} \sim
\omega^{-1.15}.
\end{eqnarray}
Due to the increase of the
mean of $X(t)$,
spectral estimation from the periodogram $X(\omega)$
(implying a periodical extension in time domain)
has to be discussed carefully.
As $\sum_{t=0}^{2^{N}-1}X(t)=3^N$, the average increase is
$\displaystyle
\langle\! X(t)\!\rangle_{\rm L}=
%
%
\langle X(t)\rangle_{\{0\ldots2^{N}-1\}}
=t^\beta$
with $\beta=\log(3/2)/\log(2) \approx 0.585$.
Further, the signal exhibits an increasing variance
$\langle X(t)^2\rangle_{\{0\ldots2^{N}-1\}} = t^\gamma$
with $\gamma=\log(5/2)/\log(2) \approx 1.32$.
Consequently,
we investigate two variants of $X(t)$:
A suitable per definitionem mean-free
sum signal defined by
$Y(t)=X(t) - (1+\beta)t^{\beta}$,
%
and a mean-free signal with non-increasing
variance
\begin{eqnarray}\label{zt}
Z(t)=Y(t) / \langle Y(t)^2\rangle^{1/2}_{\{0\ldots t\}}.
\end{eqnarray}
The spectrum of $Y(\omega)$
can be directly obtained from
eq. (\ref{xom}) and the evaluation
of the Fourier transform of $t^\beta$:
\begin{eqnarray}
Y(\omega)=X(\omega)-(1+\beta)\mathcal{F}(t^{\beta})
\end{eqnarray}
The power spectrum of the periodically extended function $t^{\beta}$
decays (for small values of the frequency) like a power law
with an exponent of approximately -2. Thus, the decay is much stronger than $X(\omega)$
and the power spectra of $Y(t)$ and $X(t)$ deviate only slightly.
Hence, it follows that $|Y(\omega)|^2\sim |X(\omega)|^2\sim \omega^{-\alpha}$.
Similarly, $|X(\omega)|^2$ also estimates $|Z(\omega)|^2$.
We now compare these results with numerically applied
discrete Fourier transformation.
The power spectrum is fitted (least-squares)
in fig.~\ref{fig_fit}
 by a  power law with exponent
 $\alpha$ about 1.11,
being in good agreement with the analytical result $\alpha\approx 1.15$
of {eq.} (\ref{result}).
If one measures a power spectrum experimentally, this may
generically be done
by observation of resonances, where the system is coupled
with a tuneable oscillator of given
frequency and finite bandwidth.
Therefore it is quite natural
to consider an averaged spectrum.
The (incommensurable) averaging procedure applied in fig.~\ref{fig_av}
smoothes the peaks at  $\omega=2^k$.
The peak amplitudes decay as the average spectrum itself.
For commensurable averaging (inset of fig.~\ref{fig_av})
the peaks at $\omega=2^k$ disappear completely.

Note that the spectra $|X(\omega)|^2$, and $|Y(\omega)|^2$,
from FFT for $T=2^{22}$ (not shown) also display $1/f^\alpha$ behaviour,
with exponents $\alpha_X=1.12$, and $\alpha_Y=1.11$ , respectively.
Moreover, we have used the method of a sliding window that normalize the fluctuations \cite{Hu2001} of the detrended signal $Y(t)$, i.e. $\tilde
Z(t)=Y(t+l)/\langle Y^2\rangle_{\{t-l+1,t+l\}}^{1/2}$. \clearpage For different values of the window width $2l$, the power spectra
exhibit power law behaviour with exponents of about $\alpha\approx 1.1$.
Thus $1/f^{\alpha}$ spectra appear to be a robust
property of Sierpinski signals.
\paragraph*{Amplitude distribution:}
Many systems exhibiting $1/f$ noise possess a Gaussian amplitude
distribution \cite{voss1978}. In this paragraph we calculate the
amplitude distribution $H_N(2^k)$ of $X(t)$ analytically where
$H_N(2^k)$ denotes the frequency occurrence of
$X=2^k,~k=0,1,\ldots$ for a signal length of $T=2^N$.
The number of $2^k$s in the signal up to $t=2^N-1$ is the number
of $2^k$s plus the number of $2^{k-1}$s in the
signal sequence up to $t=2^{N-1}-1$,
i.e. $H_N(2^k)=H_{N-1}(2^{k})+H_{N-1}(2^{k-1})$.
For the boundary condition $H_N(1)=H_1(2)=1$
we obtain a sum of binomial coefficients
$H_N(2^k)=\sum_{j=1}^N {j \choose k}={N+1 \choose k+1},~k\ge1$,
which simplifies for $N\gg1$ to 
$2^N$ times 
the Gaussian distribution:
\begin{eqnarray}\label{ampdistr}
H_N(2^{k-1})\approx\frac{2^N}{(\pi N/2)^{1/2}} e^{-\frac{(k-N/2)^2}{N/2}}.
\end{eqnarray}
Hence, the amplitude distribution of the occurrence of powers of 2
is Gaussian for fixed $N$.
The
(averaged) amplitude distributions for
$Y(t)$ and $Z(t)$ differ from $H_N(2^k)$ but possesses
a similar shape as $H_N(2^k)$.
Note that the variance distributions for 
$X(t)$, $Y(t)$, and $Z(t)$,
are not Gaussian but well defined by {eq.} (\ref{ampdistr}).

As a final point,
numerical simulations show
that the averaged signals are robust against noise, {i.e.}
initial conditions with more than a single 1.

In analogous situations in less simply defined systems, power laws
have also been observed in spatial spectra of the scum on fluid
surfaces and in the random baker map \cite{AntonsenPRL95}, and in
the temporal spectra in dissipative dynamics governed by the
Lorenz equations \cite{pikovskyPRE95} being related to the
Thue-Morse sequence
\cite{pikovskyPRE95,zaksPRL96,zaksJSP97,zaksPRE01}. In fact the
Thue-Morse dynamics $1 \rightarrow (1,-1), -1 \rightarrow (-1,1)$
itself maps on the string replication rule $u_n \rightarrow
(u_n,(-1)\cdot u_n)$ for generation of the spatial sequence. Being
not equivalent, but of striking similarity to eq. (\ref{eq_formal})
for generation of the temporal Sierpinski signal series, the
Sierpinski dynamics itself does not follow a replication rule.
While the analytic solution of the Thue-Morse sequence is
$(-1)^{\rm CrossSum(Binary(x))}$ \cite{zaksJSP97}, in analogy to eq.
(\ref{binrow}), the averaged exponents of the resulting spectra
are different.

To conclude, the one-dimensional Sierpinski automaton
generates $1/f^\alpha$ spectra in the number of active states,
and can therefore be considered as one of the simplest models
 generating $1/f^\alpha$ spectra.
While the Sierpinski automaton is rather a caricature, the
approach of studying the sum signal, or total (in-)activity, and
its spectrum, can be transferred to more realistic models and
compared directly with experiments.
Although exact Sierpinski patterns with long-range
correlations remain to be experimentally challenging,
we conjecture that $1/f^{\alpha}$ spectra in a suitable
sum signal can be identified in every
experimental setup exhibiting Sierpinski patterns.
\\[1.0ex] \indent {\sl Acknowledgements:}
We would like to thank P.J. Plath, O. Rudzick, and S. Bornholdt for fruitful discussions.
\vspace*{-3.5ex}

\end{document}